# Improving precision of objective image/video quality metrics

Majid Behzadpour[1] and Mohammad Ghanbari[*1,2]    ORCID: [0000-0002-5482-8378](#)

[1] Department of Electrical and Computer Engineering, College of Engineering, University of Tehran, Tehran, Iran, E-mails {majid.behzadpour@ut.ac.ir; ghan@ut.ac.ir}

[2] School of Computer Science and Electronic Engineering, University of Essex, Colchester, UK, CO4 3SQ, ghan@essex.ac.uk

- Corresponding author

**Abstract**

Although subjective tests are most accurate image/video quality assessment tools, they are extremely time demanding. In the past two decades, a variety of objective tools, such as SSIM, IW-SSIM, SPSIM, FSIM, etc., have been devised, that well correlate with the subjective tests results. However, the main problem with these methods is that, they do not discriminate the measured quality well enough, especially at high quality range. In this article we show how the accuracy/precision of these Image Quality Assessment (IQA) meters can be increased by mapping them into a Logistic Function (LF). The precisions are tested over a variety of image/video databases. Our experimental tests indicate while the used high-quality images can be discriminated by 23% resolution on the MOS subjective scores, discrimination resolution by the widely used IQAs are only 2%, but their mapped IQAs to Logistic Function at this quality range can be improved to 9.4%. Moreover, their precision at low to mid quality range can also be improved. At this quality range, while the discrimination resolution of MOS of the tested images is 23.2%, those of raw IQAs is nearly 8.9%, but their adapted logistic functions can lead to 17.7%, very close to that of MOS. Moreover, with the used image databases the Pearson correlation of MOS with the logistic function can be improved by 2%-20.2% as well.

## 1. Introduction

In the past two decades there have been a lot of interests in both image and video processing. This is mainly due to the explosive growth of multimedia over the internet. Currently Cisco predicts by year 2022, more than 82% of the internet traffic will be video related material [1], let alone applications in social networks, which try to retrieve mages of various kinds in the net. Considering that raw image/video demands a large volume of data to be represented properly, their compression to achieve a manageable storage and transmission rate is inevitable. This is only possible at the cost of induced distortions in the processed image/video. It is highly desired to measure such distortions, by any objective measuring tool.

Considering that the ultimate receptor of visual content is the human visual system (HVS), the best and most accurate measuring device for assessing processed image/video distortions is again based on HVS. This is normally carried out by subjective tests, where a group of viewers watch a set of distorted image/video contents, and viewers' mean opinion score (MOS) is taken as the best representative of visual quality. However, this process apart from being time consuming, it requires certain laboratory set ups, which may not be feasible for all users.

To resolve MOS limitations, historically image/video quality is measured based on the difference between their unprocessed and processed versions and presented in terms of Peak-Signal-to-Noise Ratio, PSNR. However, it may be argued that PSNR is not a valid quality measure. For instance, if the original non-distorted image is shifted even by one pixel, the difference between the original signal and its shifted version can show a significant drop in PSNR, whereas the shifted image quality is subjectively perfect. Or, if image dimensions are arbitrarily enlarged/reduced such quality measure is useless. Moreover, PSNR value is not an indication of absolute acceptable video quality, nor it can be used to compare two different visual contents. Despite this, PSNR is a valid criterion, in comparing images/video of the same content, provided their dimensions are not altered. In [2], it is shown that if image content remains unaltered, improving PSNR can definitely improve MOS. This is the reason all video codecs, through rate-distortion optimization try to minimize coding distortion (maximize PSNR) for the best subjective quality.

Over the past two decades a number of image quality assessment (IQA) tools have been devised that well alleviate PSNR limitations. However, a common problem with all these devices is that, they lose precision and accuracy at high image quality range. The main contribution of this paper is to devise a Logistic Function (LF) to improve the performance of these quality metrics. Through the experiments we show how LF can be easily added to all of these measuring tools, not only to improve their precision but also to increase their correlations to MOS.

The rest of the paper is organized in the following order. Section 2 looks at some of the most common IQA measuring tools and their common limitations. Section 3 introduces the proposed Logistic Function (LF), and through experiments show LF can increase the Pearson Correlation of all IQAs with the MOS. Section 4 extends the proposed method to enhance the widely used Video Multimethod Assessment Fusion (VMAF) of measuring video quality. Finally, Section 5 draws some concluding remarks.

## 2. Popular Image Quality Assessment (IQA) tools

PSNR is a kind of full reference model, but it is sensitive to spatial positions of either reference or processed image. A family of full reference meters that do not have such sensitivity, are based on structural similarity, the so-called structural similarity index [3]. In this method as long as processing does not alter the structure of neighboring pixels, human visual system is not sensitive to any added distortion due to processing. Another words, it is assumed that the human visual system is sensitive to changes in the local structure. In the past two decades, numerous methods based on structural similarity have been devised.

For instance, in multi scale structural similarity [4] it is assumed that the human visual system adapts itself to extract structural information of the scene, and hence structural similarity can provide a good measure of perceived image quality. Weighting structural similarity for better adaptation is presented in IWSSIM [5]. In [6] Local weight is calculated based on the symmetry model of the reference image and more weights are given to certain areas. Using such criterion, [7] introduces VSI, a visual saliency-induced index for perceptual image quality assessment, where more weight is given in the pooling strategy. FSIM: a Feature Similarity Index for image quality assessment is described in [8]. Since human visual system is more sensitive to image edges, FSIM is mainly an edge-sensitive image quality assessor. The super-pixel method, known as SPSSIM, is another well-known and new model that divides images into meaningful areas and the evaluation model is based on the local quality of these areas [9]. Finally, an image quality assessment method based on edge-feature image segmentation (EFS) was proposed in [10].

Although these variants of Image Quality Assessment (IQA) methods have some gains or deficiencies over each other, they all suffer from a common deficiency that, at high image quality range they tend to saturate. This makes their measured values at high image quality to lose accuracy and make them almost unreliable quality meter. Figure 1 shows relation between MOS and the objective quality scores by some of these methods for a TID2013 [16] image database. They include: SSIM [3], IW-SSIM [5], MS-SSIM [4], VSI [7], FSIM [8], FSIMc [8], SPSIM [9], EFS [10] and GMSD [11]. As seen at high image quality, all the measured values saturate and lose precision. At lower image quality range, although some behave better than SSIM, but still images at this range are scattered. This paper aims to alleviate these shortfalls and hopefully to improve their correlations with the MOS.

Before explaining how the precision of measurement and its correlation with MOS can be improved, let us briefly explain how each of these measures, define structure in their definitions:

(a) SSIM: The structural similarity index measure (SSIM). which is used for measuring the structural similarity between two blocks of pixels.

(b) IW-SSIM: Information Content Weighted Structural Similarity Index for IQA, which gives extra weight to the content during pooling.

(c) MS-SSIM: Calculates the multi-scale structural similarity (MS-SSIM). This function calculates the SSIM index of several versions of the image at various scales.

(d) VSI: A Visual Saliency-Induced Index for IQA. Visual saliency (VS), puts emphasis on areas of an image which will attract the most attention of the human visual system.

(e) FSIM: A Feature Similarity Index for IQA. It is based on the fact that human visual system (HVS) understands an image mainly according to its low-level features. Specifically, the phase congruency (PC), which is a dimensionless measure of the significance of a local structure.

(f) FSIMc: is a FSIM which also uses colour components information in its calculation.

(g) SPSIM: A Superpixel-Based Similarity Index for IQA. It is based on the fact that a superpixel is a set of image pixels that share similar visual characteristics and is thus perceptually meaningful.

(h) EFS: An image quality assessment method by edge-feature-based image segmentation (EFS).

(i) GMSD: Gradient Magnitude Similarity Deviation for IQA. The reason for using such a measure in structural similarity is that: image gradients are sensitive to image distortions, while different local structures in a distorted image suffer different degrees of degradations.

On the significance of the above measures, it is worth noting that Wang, Bovic and Lu in a highly cited article [20] have answered the question of "why image quality assessment is so difficult", and they have concluded that a correct way in measurement is to model the image degradation as a structural distortion instead of error (which is used in PSNR). This is a good indication of why structurally based distortion measure is so popular.

In a recent survey paper [21], the suitability of 13 different SSIM-based IQA methods, as well as the PSNR were tested in measuring the quality of error-concealed video clips. The aim was to find out which of these methods best measure the quality improvement of lost packetized video clips. In these tests, the quality of error-concealed video frames alone as well as the whole video clips were evaluated by these 13 well-known I/VQA metrics It was generally concluded that, compared to conventional video quality assessment, the metrics used in this study were not always as successful, and also their relative performance was different.

However, all these methods have the common weakness that, their discrimination of quality, particularly at high image quality range is poor and we hope by alleviating this deficiency, the usage of structural similarity based image quality meters can even become more widely spread.

### 3. Proposed Logistic Function (LF)

Although as Figure 1 shows, the variants of SSIM family improve the shortfall of SSIM at mid to low image quality range, but at high image quality, all of them similar to SSIM suffer from precision and accuracy. In [12] problem of SSIM has been mathematically studied and some improvements on image SSIM-IQA has been reported. However, in this paper we mainly look at on loss of precision of these metrics at high image quality. We also show that how a Logistic Function (LF) can be derived to improve such shortcoming of these meters and in particular it is extended to video quality measurement.

Considering the saturated nature of MOS at high IQA of these methods, it appears if the IQA values at high quality are exponentially modified, such that a small increment in IQA, can make a distinct difference on MOS, then these shortfalls are enhanced. For example, if we define Logistic Function (LF), as given in Equ 1:

$$LF = 1 - \sqrt{1 - IQA(x \times y)} \qquad (1)$$

It can map a dense variation at high value of input to a linear output. Figure 2 shows the mapping function of IQA of objective metric VSI [3] to a LF-VSI modified objective metric.

To explain how LF function defined in Equ 1 can alleviate saturation problem and hence improve precision of SSIM based measured values, consider the crossed points × on the curve of Fig 2. In this figure the objective score (or MOS) on the horizontal axis is defined in the range of 0-1. At high values of MOS or IQA, according to Equ 1, 1- IQA becomes very small. Taking square root of smaller values, leads to larger outputs. That is; the smaller is the value, the larger becomes its square root, and hence separating these points wider apart from each other. On the other hand, at lower values of IQA, 1-IQA gets larger and its square root becomes closer to each other. Thus, such definition of LF makes larger values of IQA to be separated from each other and smaller values get closer to each other. However, SSIM measured values at low values are sufficiently separated from each other, such that this kind of closeness does not impair its precision at lower values.

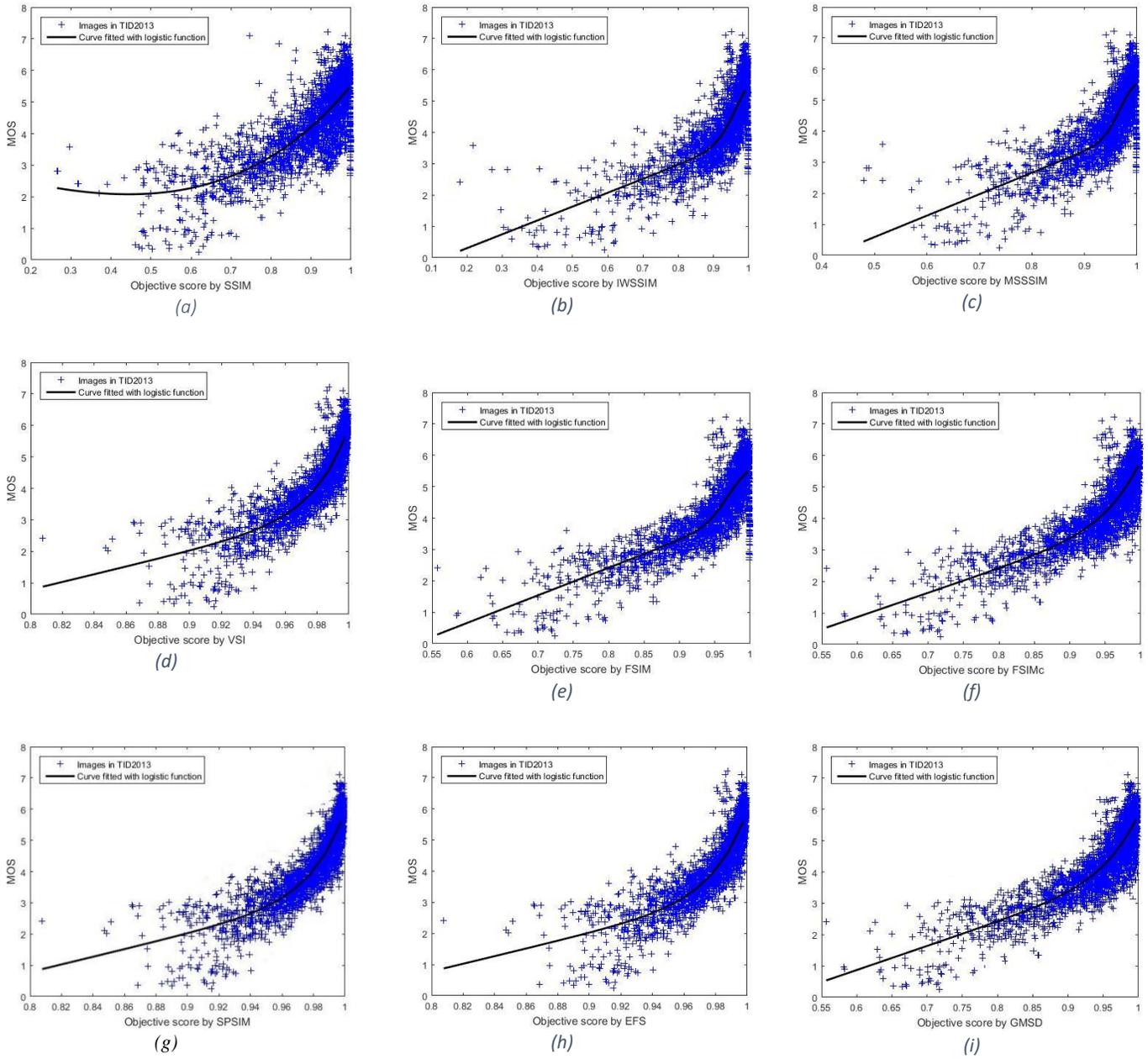

Fig. 1. Scatter plots of subjective MOS against scores obtained by model prediction on the TID2013 database. (a) SSIM, (b) IW-SSIM, (c) MS-SSIM, (d) VSI, (e) FSIM, (f) FSIMc (g) SPSIM, (h) EFS (i) GMSD.

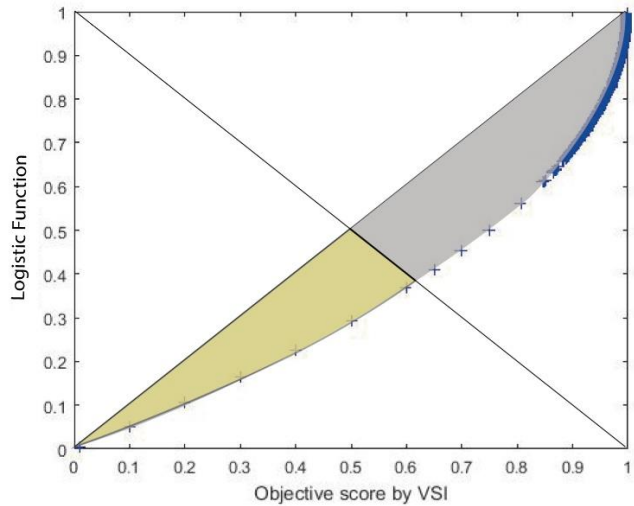

Fig. 2. Objective score VS Logistic Function (LF)

It is worth mentioning that, the logistic function (LF) of Equ (1) can be incorporated within each measuring device, to directly calculate LF value, rather than the IQA value. Please note that in quality assessment tools, normally for each block of pixels, or segment of images an IQA is calculated and the aggregate of IQAs of blocks/segments represent the final IQA score of an image. Alternatively, one may just use the overall output of IQA and map it to LF value as the final score. However, since Equ 1 is not a linear function of IQA, the two methods are not exactly equivalent, and the former has a better accuracy over the latter. However, for simplicity and being more conservative, we have taken the worst case of mapping the final IQA to LF value. This is executed throughout all experiments and had we used per block/segment LF values and their sum are amalgamated into a final LF, the results would have been better.

To see how the defined Logistic Function (LF) can improve the saturated high image quality, the IQA values measured by the used SSIM [3], IW-SSIM [5], MS-SSIM [4], VSI [7], FSIM [8], FSIMc [8], SPSIM [9], EFS [10] and GMSD [11] methods are mapped to LF, and shown in Figure 3. In the graphs of Fig. 3, after measuring the IQA by each measuring device, the derived IQA is mapped to its equivalent LF using Equ (1), and hence the MOS is drawn with respect to their LF values.

As seen in Figure 3, this time MOS has a better linear relation to the LF versions of the used IQA methods. Moreover, the resultant quality measure can have a higher correlation to the MOS too. Table 1 shows the Pearson correlation coefficient PLCC of 9 structure similarity based IQAs for 4 sets of image databases of: CISQ [13], LIVE [14], tid2008 [15] and tid2013 [16]. In this Table correlation between the MOS and measured IQA value of each method with and without their Logistic Functions are tabulated. The Table shows that for every measuring device, its LF version has much better correlation to MOS than IQA of that measure itself.

Please note that at the very high values of quality of Fig1, there are drops in quality and these are also present in their LF-versions in Fig 3. This is due to the fact that some of the images are in colour and are subjectively rated with colour fidelity, but in the objective measure, only luminance components are calculated (or vice versa, if they are in black and white, colour comments are also included in the objective measure). For instance, by comparing the scatter diagram of FSIM (without colour) with that of FSIMc (with colour), where the objective measure also includes colour components, the difference in dropping the quality at high end can be verified.

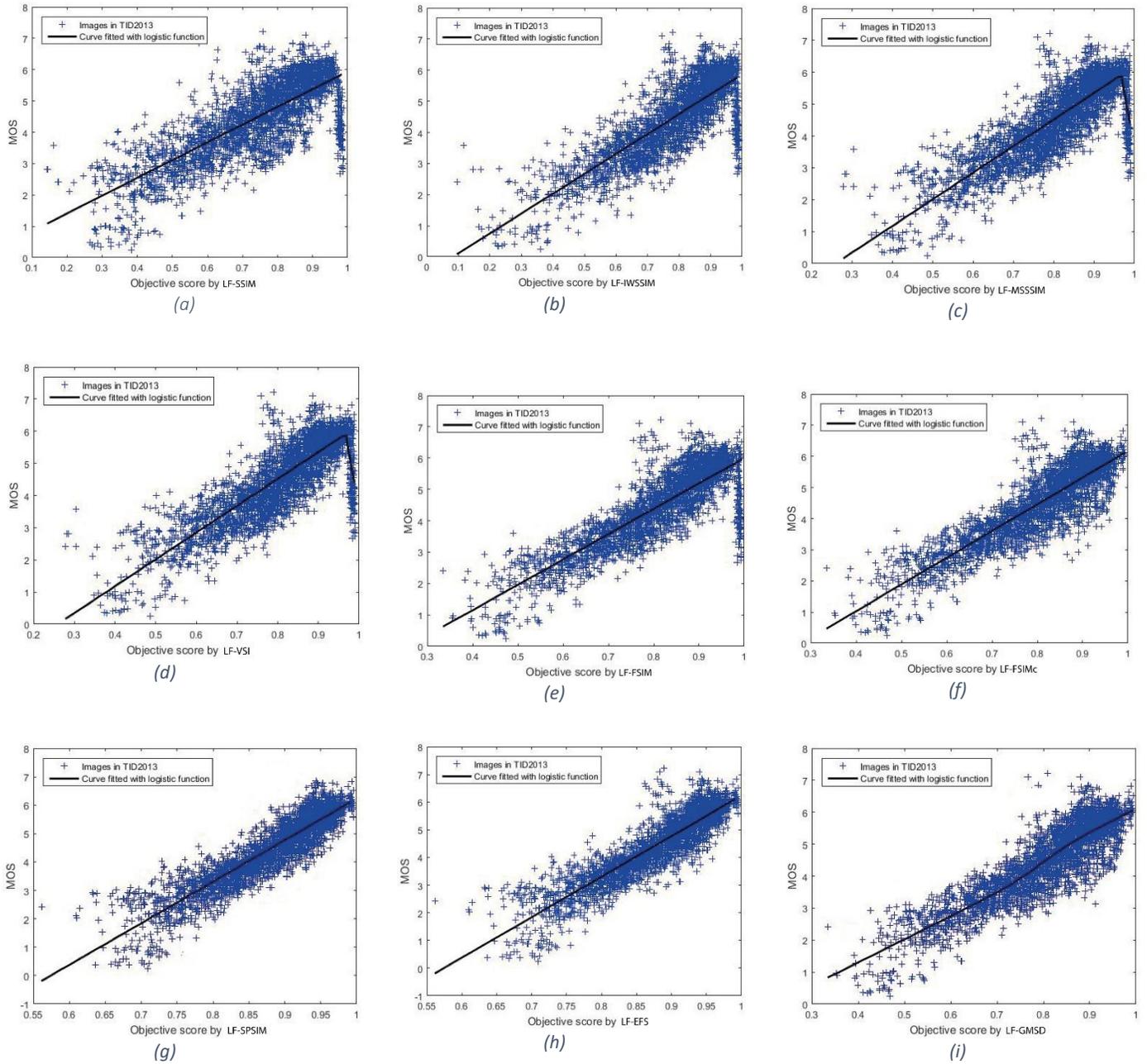

Fig. 3. Scatter plots of subjective MOS against scores obtained by model prediction with Logistic Function (LF) on the TID2013 database. (a) **LF**-SSIM, (b) **LF**-IW-SSIM, (c) **LF**-MS-SSIM, (d) **LF**-VSI, (e) **LF**-FSIM, (f)**LF**-FSIMc, (g) **LF**-SPSIM, (h) **LF**-EFS, (i) **LF**-GMSD.

Another important point to note is that, if among the SSIM family, a method performs better than the other, its LF mapped version will also perform better. The reason is that, according to Equ (1), each LF version of measured IQA is directly related to its IQA value. For instance, in Table 1, among the 9 tested IQAs, if GMSD is the best measuring device for CISQ image database, its LF version also has the highest performance among the all LF versions for this sequence. One may inspect all image databases of Table 1, for such property. This implies that, the IQA value of any measuring device can be mapped to LF, to improve its precision without damaging its correlation accuracy to MOS. The more significant point is that , since with the defined LF, higher quality points are better separated from each other and lower quality ones come closer (as explained in the text on page 4, before Fig 2), by bringing the measured quality values to their correct positions, the correlation between LF adapted SSIM with MOS increases. For instance, in all images and databases listed in Table 1, the Pearson Correlation between MOS and LF can be 2%-20.2% better than the correlation between MOS and measured IQA itself.

Table 1. Comparison between PLCC of MOS of various IQAs without and with their Logistic Functions (LF) for 4 image datasets

|  | CISQ | | LIVE | | tid2008 | | tid2013 | |
|---|---|---|---|---|---|---|---|---|
|  | IQA | LF | IQA | LF | IQA | LF | IQA | LF |
| ssim | 0.7916 | **0.8526** | 0.829 | **0.922** | 0.7401 | **0.7732** | 0.7596 | **0.7746** |
| iw-ssim | 0.7947 | **0.8891** | 0.8029 | **0.9207** | 0.8086 | **0.8555** | 0.7638 | **0.8172** |
| FSIM | 0.8048 | **0.893** | 0.8586 | **0.9407** | 0.8301 | **0.8724** | 0.8195 | **0.8408** |
| FSIMc | 0.8208 | **0.9032** | 0.8595 | **0.9472** | 0.8341 | **0.8747** | 0.8322 | **0.8749** |
| Ms-ssim | 0.772 | **0.87** | 0.767 | **0.9035** | 0.7897 | **0.8406** | 0.7773 | **0.8145** |
| VSI | 0.8392 | **0.9154** | 0.7647 | **0.9035** | 0.8107 | **0.8681** | 0.8373 | **0.8957** |
| SPSIM | 0.8583 | **0.9344** | 0.7985 | **0.9599** | 0.8312 | **0.8929** | 0.8468 | **0.9092** |
| GMSD | 0.8812 | **0.9542** | 0.8629 | **0.9602** | 0.812 | **0.8717** | 0.8031 | **0.8542** |
| EFS | 0.8412 | **0.9215** | 0.8591 | **0.9445** | 0.8276 | **0.8801** | 0.8431 | **0.9011** |

Apart from higher PLCC of LF measured metrics, they have a better precision, not only at high quality, but also at medium and low quality as well.

To investigate the precision of the Logistic Function across all image quality ranges, as well as those in the structurally based similarity measures, we borough the idea of image quality analysis method in large databases used in [16]. In image analysis defined in [16], the MOSs of about 3000 images in the tid2013 image database are classified into three groups based on their quality range, each one of nearly 1000 images. First of all, the whole images are rated into 8 ranges (0-8). The first group called "bad quality" have an MOS in the range of 0.242-3.94. The second class called "middle quality" group contains images with MOS in the range of 3.94-5.25. Finally, the third group contains "good quality" images with MOS higher than 5.25.

Since our goal is to measure precision, we group the images in a known value of fixed quality of 2, 4 and 6 for bad, middle and good quality respectively. First of all, we do not need to take 1000 images, instead we have taken only 10 images for each group. We will show that even such a small sample, can prove our concept of precision measure.

Table 2a shows SSIM values and their LF versions along with MOS for 10 images, selected from the tid2013 database at almost high MOS score of 6 (good quality). Similarly, Tables 2b and 2c show these values for 10 images of middle and bad quality, respectively in the MOS scores of 4 and 2. The Tables also include the averages of MOS for the three different quality, as well as averages of their

SSIM and LF versions of SSIM (LF-SSIM). Inspection of these data reveal the following interesting outcomes:

1- The difference between the average MOS of good quality video from the average of middle quality is 6.3487- 4.5120= 1.8367. Considering that the MOS range is 0 to 8, then this difference indicates a precision of 0.2296, equivalent to almost 23% difference in quality. Note that theoretically, the difference between good quality of 6 and middle quality of 4 (if all had MOS of exactly 6 and 4), is 0.25, corresponding to 25%, not much different from 23%. Thus, even a small sample of 10 images, show such a high precision on MOS. However, such a difference on the average of SSIM is only 0.9958 – 0.9757= 0.0201. Meaning the SSIM precision in discriminating a good image quality from a middle quality is only 2%. This is significantly less than 23% of MOS and shows its weakness in assessing video quality at high IQA range. On the other hand, the difference between LF-SSIM from good to middle quality is 0.9385-0.8449=0.0936. This is equivalent to nearly 9.4%, almost 4.7 times better precision than SSIM at this quality range.

2- The proposed method not only improves precision at high quality range, as shown above, it also has a better performance at middle and bad quality ranges as well. This can be investigated by looking at the average values of MOS, IQA of SSIM and its LF version (LF-SSIM), in going from middle to bad quality. In this case for MOS, the difference between them is: 4.5120-2.6594= 1.8526. This on 0 -8 MOS scale is equivalent to 1.8526/8=0.2316, which is 23.16% precision (again very close to the theoretical value of 25%). Such a precision on the SSIM discrimination between middle and bad quality is: 0.9759-0.8865=0.0894, equivalent to 8.9% precision. However, this precision for LF-SSIM is: 0.8449-0.6676= 0.1773. This means the precision of LF version is 17.73%, which is much closer to MOS discrimination value of 23.16% than the SSIM alone of 8.9%. It is almost twice the precision of SSIM at bad-middle quality range.

We have tested the above scenarios with all the SSIM family measuring tools. They showed almost the same behavior as was explained above on SSIM.

Table 2a: Values of each IQA metric and their LF-SSIM along with MOS for good quality images from database tid2013 [16]

| image | MOS | SSIM | LF-SSIM |
|---|---|---|---|
| 'i01_02_2.bmp' | 6.10811 | 0.994849 | **0.9279** |
| 'i03_08_1.bmp' | 6.34211 | 0.997039 | **0.9453** |
| 'i04_02_1.bmp' | 6.275 | 0.992736 | **0.9148** |
| 'i05_16_1.bmp' | 6.15 | 0.993907 | **0.9219** |
| 'i07_09_1.bmp' | 6.42222 | 0.990429 | **0.9021** |
| 'i23_16_1.bmp' | 6.33333 | 0.997993 | **0.9542** |
| 'i03_16_1.bmp' | 6.82051 | 0.9969 | **0.9444** |
| 'i12_16_1.bmp' | 6.52632 | 0.998817 | **0.9654** |
| 'i24_02_1.bmp' | 6.44444 | 0.996225 | **0.9384** |
| 'i08_04_1.bmp' | 6.06452 | 0.99915 | **0.9709** |
| **avg** | **6.348656** | **0.995805** | **0.93853** |

Table 2b: Values of each IQA metric and their LF-SSIM along with MOS for middle quality images from database tid2013 [16].

| image | MOS | SSIM | LF-SSIM |
|---|---|---|---|
| 'i01_02_4.bmp' | 4.71429 | 0.980467 | **0.86** |
| 'i01_17_3.bmp' | 4.08108 | 0.963197 | **0.8082** |
| 'i03_04_4.bmp' | 4.6 | 0.978184 | **0.8521** |
| 'i03_08_3.bmp' | 4.47368 | 0.970649 | **0.8286** |
| 'i04_17_3.bmp' | 4.78049 | 0.974127 | **0.8391** |
| 'i05_01_3.bmp' | 4.84615 | 0.979057 | **0.8551** |
| 'i05_02_5.bmp' | 4.25641 | 0.978483 | **0.8531** |
| 'i05_19_3.bmp' | 4.225 | 0.983719 | **0.8724** |
| 'i07_06_2.bmp' | 4.61364 | 0.973127 | **0.836** |
| 'i23_04_5.bmp' | 4.52941 | 0.975796 | **0.8442** |
| **avg** | **4.512015** | **0.975681** | **0.84488** |

Table 2c: Values of each IQA metric and their LF-SSIM along with MOS for bad quality images from database tid2013 [16].

| image | MOS | SSIM | LF-SSIM |
|---|---|---|---|
| 'i03_09_5.bmp' | 2.47368 | 0.809075 | **0.563** |
| 'i20_10_5.bmp' | 2.74359 | 0.871029 | **0.6409** |
| 'i09_07_5.bmp' | 2.70968 | 0.848929 | **0.6113** |
| 'i06_07_5.bmp' | 2.68571 | 0.887359 | **0.6643** |
| 'i25_10_4.bmp' | 2.14706 | 0.928 | **0.7317** |
| 'i08_07_5.bmp' | 2.3871 | 0.883728 | **0.659** |
| 'i13_15_1.bmp' | 2.73171 | 0.945102 | **0.7657** |
| 'i18_22_4.bmp' | 2.95238 | 0.912675 | **0.7044** |
| 'i08_03_4.bmp' | 2.96875 | 0.877367 | **0.6498** |
| 'i25_22_5.bmp' | 2.79412 | 0.901483 | **0.686** |
| **avg** | **2.659378** | **0.886475** | **0.66761** |

The above analysis, indicates that, discrimination of Logistic Function version of SSIM not only is more than 4 times accurate than SSIM itself at good image quality range, but its precision at middle to bad quality is still twice better.

It is worth noting that the used Logistic Function (LF) in Equ 1, can be defined in a variety of ways. For instance, we tested, Equs 2 and 3 as:

$$LF = 1 - \sqrt[2]{1 - IQA^2} \qquad (2)$$

$$LF = 1 - \sqrt[3]{1 - IQA^2} \qquad (3)$$

And various variants of these equations. The outcome of these investigation was that, some may discriminate good from middle quality at higher precision and become very close to MOS precision, but their middle to bad quality discrimination range deteriorates. It appears LF defined in Equ 1 is the best compromise in discriminating good from middle and middle from bad quality average values.

## 4. Logistic Function (LF) for VMAF, in video quality assessment

Image quality assessment IQA parameters can also be used to measure video quality, since video is made up of a series of video frames/pictures. For instance, in [17], it is reported that for a 10 second video, average of 20% of worst IQAs measured on frame-by-frame bases has a very high correlation with the subjective scores. However, since the advent of Video Multimethod Assessment Fusion (VMAF), that predicts subjective video quality by diffusing multiple quality metrics into a single score through machine learning optimization, this method has become the most popular video quality meter. It was originally collaboratively devised by researchers in Netflix and colleagues of Professor CC Kuo, at the university of Southern California [18]. However, over the years, through more works and tests like deep learning and better training, it has become a de facto method for video quality assessment, and almost everyone uses it [19].

It would be interesting to see if the used logistic function (LF) for image quality meters, can also improve VMAF's performance for video. We have tested Netflix database, comprising of 75 video sequences and the LIVE video of 150 video sequences. The scatter diagram of these sequences, for VMAF and its LF version (LF-VMAF, which uses VMAF metric in Equ 1) are shown in Fig. 4. In these tests, while PLCC of VMAF for Netflix sequence was 0.965, this value with LF-VMAF was 0.946 (0.02 point worse). On the Live video, while PLCC of VMAF to MOS was 0.7549, this value with LF-VAMF was 0.7704 (almost 0.02 better).

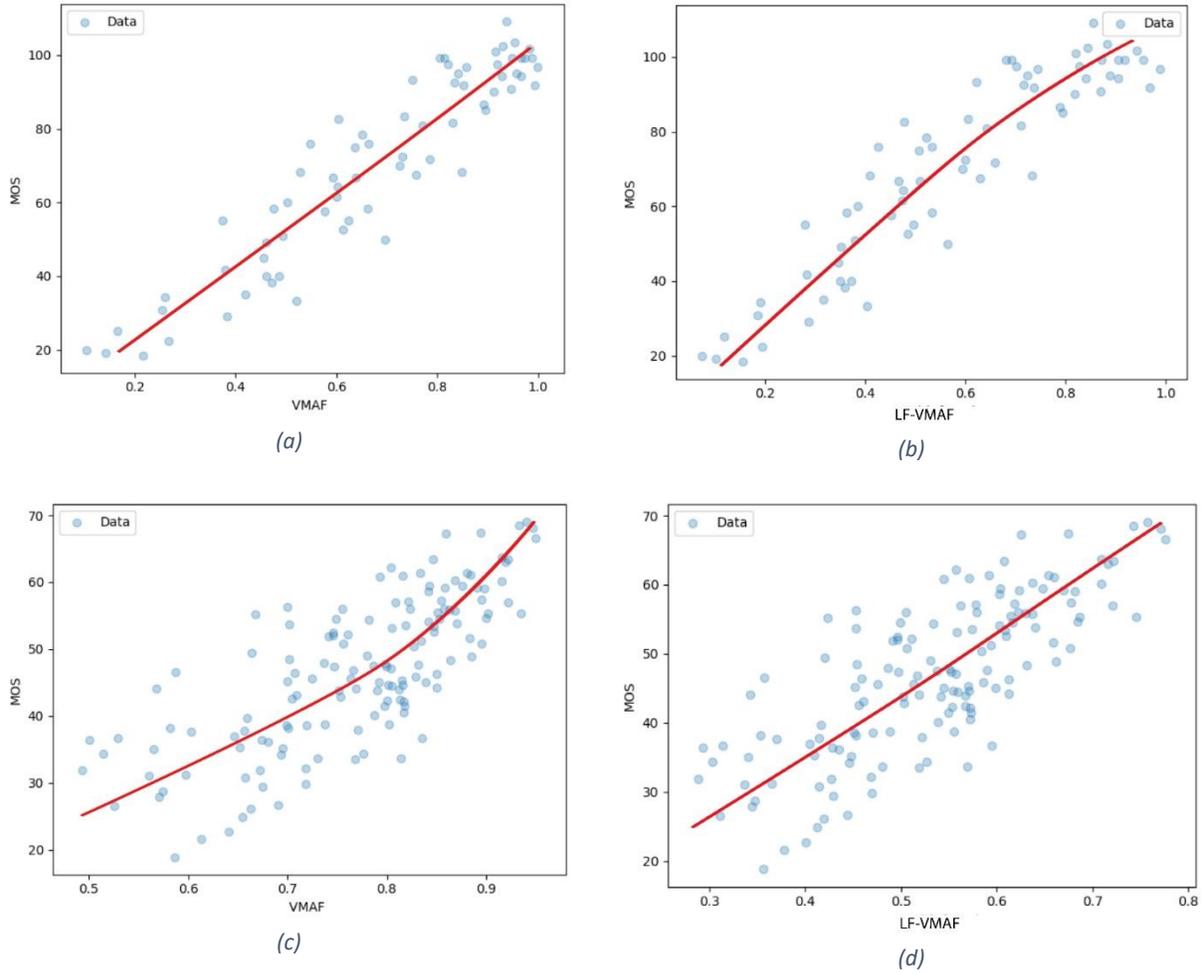

Fig. 4. Scatter plots of subjective MOS against scores, Netflix database (a) VMAF, (b) **LF**-VMAF, and Live database (c)VMAF, (d) **LF**-VMAF.

It is important to note that, these scatter diagrams, compared with those of images, shown in Figs 1 and 3, that contain more than 1000+ images (e.g. tid2013 has 3000 images), are very sparse. These sparsely scattered points do not have enough data to show the saturation limitation of video quality metrics. Had we had in the order of 1000 video clips we could have better results with LF-VMAF. This can be verified by comparing LIVE and Netflix datasets, where they have respectively 150 and 75 video sequences, where the dataset with larger video sequences has a better performance with LF-VMAF. It should be noted that testing with larger database of video sequences is labor intensive, as 150 video sequences of LIVE video required 40GB storage.

## 5. Conclusion

Image quality assessment (IQA) tools are widely used in evaluating quality of processed images. They belong to a family of structural similarity index (SSIM) method, that correlate well with the human visual systems behavior. Through more than two decades, numerous versions of SSIM-based image quality assessment meters have been devised. Their test results show some improvements of one method over the other. However, they all suffer from loss of precision, especially at high image quality range.

In this paper we have shown that a simple logistic function can be added to the outcome of these measuring devises, to improve their precisions. Throughout the experiments we have shown that, the added logistic function not only improves precision at high quality images, those of low-quality ones can also be improved. This improvement in precision can also increase the Pearson Correlation of the objective measures with the mean opinion scores (MOSs). For all images of databases listed in Table 1, while Pearson correlation of MOS to LF of any measured device has a minimum improvement of 2%, its maximum improvement is as high as 20.2%.

For analysis of a large database of images, if we divide them into three groups of bad, middle and good quality images, while the MOS of good-quality images, had almost 23% precision, this value for IQAs at this quality was only 2%, but their adapted logistic function at this quality was 9.4%. Such modification could also improve measured quality precision at middle to bad quality. In this case, while precision of MOS at these quality ranges was 23.2%, that of raw IQA was 8.9% and their logistic function version was increased to 17.7%, very close to the MOS range.

Finally, we have tested the impact of defined logistic function on video quality meter, especially the widely used VMAF. Although a limited number of video sequences were tested, but their outcomes indicate that a logistic function can also improv the precision of video quality meters too.

## References

1. Cisco: Cisco visual networking index: Forecast and methodology, 2017{2022 (White Paper) (2019)
2. Huynh-Thu, Q., Ghanbari, M. (2008) Scope of validity of psnr in image/video quality assessment. *Electronics Letters* 44(13), pp800–801
3. Z. Wang, A. C. Bovik, and E. P. Simoncelli, (2004) Image quality assessment: from error visibility to structural similarity, *IEEE Transactions on Image Processing*, vol. 13, pp. 600-612
4. Z. Wang, E. P. Simoncelli, and A. C. Bovik, (2003) Multiscale structural similarity for image quality assessment, IEEE *Conference Record of the Thirty-Seventh Asilomar Conference on Signals, Systems and Computers,* vol. 2.
5. Z. Wang, and Q. Li, (2011) Information content weighting for perceptual image quality assessment, *IEEE Transactions on Image Processing,* vol. 20, no. 5, pp. 1185-1198.
6. W. Zhang, A. Borji, Z. Wang, P. Le Callet, and H. Liu, (2016) The application of visual saliency models in objective image quality assessment: A statistical evaluation, *IEEE Trans. Neural Netw. Learn. Syst.*, vol. 27, no. 6, pp. 1266–1278.
7. L. Zhang, Y. Shen, and H. Li, (2014) VSI: A visual saliency-induced index for perceptual image quality assessment, *IEEE TransActions on Image Processing*, vol. 23, no. 10, pp. 4270–4281.
8. Lin Zhang, Lei Zhang, X. Mou and D. Zhang, (2011) FSIM: A Feature Similarity Index for Image Quality Assessment," *IEEE Transactions on Image Processing*, vol. 20, no. 8, pp. 2378-2386.
9. W. Sun, Q. Liao, J.-H. Xue, and F. Zhou, (2018) SPSIM: A superpixel-based similarity index for full-reference image quality assessment, *IEEE TransActions on Image Processing*, vol. 27, no. 9, pp. 4232–4244.

10. ShiZaifeng, ZhangJiaping, CaoQingjie, PangKe, LuoTao, (2018) Full-reference image quality assessment based on image segmentation with edge feature, *Signal Processing* vol. 145, pp. 99-105, 2018

11. W. Xue, L. Zhang, X. Mou, and A. Bovik, (2014) Gradient magnitude similarity deviation: A highly efficient perceptual image quality index, *IEEE Transactions on Image Processing*, 23(2):684–695.

12. D. Brunet, E. R. Vrscay, and Z. Wang, (2012) On the mathematical properties of the structural similarity index," *IEEE Transactions on Image Processing*, vol. 21, no 4, pp. 1488–1499.

13. E. C. Larson and D. M. Chandler, Categorical Image Quality (CSIQ) Database 2009 [Online]. Available: http://vision.okstate.edu/csiq

14. K. Seshadrinathan, R. Soundararajan, A. C. Bovik and L. K. Cormack, (2010) Study of Subjective and Objective Quality Assessment of Video", *IEEE Transactions on Image Processing*, vol.19, no.6, pp.1427-1441.

15. N. Ponomarenko, V. Lukin, A. Zelensky, K. Egiazarian, M. Carli, and F. Battisti, (2009) TID2008—A database for evaluation of full-reference visual quality assessment metrics," *Adv. Modern Radioelectron.*, vol. 10, pp. 30–45.

16. N. Ponomarenko, L. Jin, O. Ieremeiev, V. Lukin, K. Egiazarian, J. Astola, B. Vozel, K. Chehdi, M. Carli, F. Battisti, C.C.J. Kuo (2015) Image database tid2013: Peculiarities, results and perspectives. *Signal Processing: Image Communication*, 2, 3, 4

17. KT Tan, M Ghanbari, (2000) A multi-metric objective picture-quality measurement model for MPEG video, *IEEE Transactions on Circuits and Systems for Video Technology* 10 (7), 1208-1213.

18. TJ Liu, W Lin, CCJ Kuo, (2012) Image quality assessment using multi-method fusion'*, IEEE Transactions on Image Processing* 22 (5), 1793-1807.

19. VMAF: Perceptual video quality assessment based on multi-method fusion, Netflix, Inc., 2017-07-14, retrieved 2017-07-15

20. Z. Wang, A. C. Bovik and L. Lu, (2002) "Why is image quality assessment so difficult?," *2002 IEEE International Conference on Acoustics, Speech, and Signal Processing*, Orlando, FL, 2002, pp. IV-3313-IV-3316, doi: 10.1109/ICASSP.2002.5745362.

21. M. Kazemi, M. Ghanbari and S. Shirmohammadi, (2020)"The Performance of Quality Metrics in Assessing Error-Concealed Video Quality," in *IEEE Transactions on Image Processing*, vol. 29, pp. 5937-5952, doi: 10.1109/TIP.2020.2984356.